\documentclass[11pt,a4paper]{article}
\usepackage{jheppub_kim}
\usepackage{longtable}

\usepackage{epsfig}
\usepackage{amsmath}
\usepackage{graphicx}
 \usepackage{graphics}
\usepackage[hang,nooneline,scriptsize]{subfigure}

\usepackage{array}
\begin{document}

\title{Exploring Non-perturbative Corrections in Thermodynamics of Static Dirty Black Holes}

\author[a]{Saheb Soroushfar,} 
\author[b]{Behnam Pourhassan,}
\author[c]{\.{I}zzet Sakall{\i},}
\affiliation[a]{Department of Physics, College of Sciences, Yasouj University, 75918-74934, Yasouj, Iran.}
\affiliation[b] {School of Physics, Damghan University, Damghan, 3671641167, Iran.}
\affiliation[c] {Physics Department, Eastern Mediterranean University, Famagusta 99628, North Cyprus via Mersin 10, Turkey.}

\emailAdd{soroush@yu.ac.ir}
\emailAdd{b.pourhassan@du.ac.ir}
\emailAdd{izzet.sakalli@emu.edu.tr}

\abstract{This study presents an investigation into the thermodynamic properties of a dirty black hole immersed in a uniform electric field within the framework of the Einstein-Nonlinear Electrodynamics (ENE)-dilaton theory. The analysis delves into various thermodynamic aspects, including heat capacity, Helmholtz free energy, and internal energy, providing insights into the behavior of the black hole under the influence of the electric field. Furthermore, the article explores the intricate interplay between quantum effects and thermodynamic behavior through the examination of quantum-corrected entropy. The study aims to shed light on the non-perturbative corrections that arise in this complex system, offering a comprehensive understanding of the modified thermodynamics of dirty black holes within the specified theoretical framework.}


\maketitle

\section{Introduction} \label{sec1}
The field of black hole research has captivated and proven vital to the realm of theoretical physics and astrophysics \cite{Wald:1984rg,Frolov:1998wf}. Black holes are enigmatic objects that form when massive stars collapse under their gravity, creating a region in space where gravity is so strong that nothing, not even light, can escape its pull. These mysterious entities originate from the implosion of colossal stars and showcase extraordinary thermodynamic traits. The groundbreaking endeavors of Bekenstein and Hawking \cite{5-0-B,8,9,HawCMP} established the bedrock principles of black hole thermodynamics \cite{Gibbons:1977mu}, revealing that these entities possess temperature, entropy, and energy resembling those of thermodynamic systems, similar to ordinary thermodynamic systems. This idea was revolutionary because it treated black holes as thermodynamic objects rather than just gravitational entities. Thus, it is understood that black holes emit thermal radiation, which is known as Hawking radiation \cite{Hawking:1982dh,Parikh:1999mf,Hawking:1976de,Sakalli:2017ewb,Sakalli:2014sea,Sakalli:2012zy,Sakalli:2016aia}. The temperature of this radiation, called Hawking temperature, is inversely proportional to the black hole's mass in the Schwarzschild family \cite{Hawking:1974rv}. This temperature is incredibly low for massive black holes, making it hard to detect in practice. However, it has significant implications for the understanding of the universe's evolution and the connection between general relativity and quantum mechanics.  It is also worth noting that especially in the non-asymptotically flat black holes, the temperature may remain constant, as being independent of the mass throughout the Hawking radiation. Such a phenomenon can occur during an isothermal process \cite{Mazharimousavi:2009nc,Pasaoglu:2009te,Sakalli:2022xrb}.\\
Entropy, another important thermodynamic property, measures the disorder or randomness present in a system. As previously mentioned, Bekenstein and Hawking \cite{Bekenstein:2008smd} discovered that black holes have an entropy directthermal fluctuations are occurringather than their volume \cite{Strominger:1996sh,Strominger:1997eq}. This finding was revolutionary because it suggested a link between the macroscopic properties of black holes and the microscopic world of quantum mechanics \cite{Li:2003vu}. The field of black hole thermodynamics has expanded our understanding of the fundamental laws of physics. It has led to significant advancements in topics such as the holographic principle, quantum gravity, and the information paradox \cite{Raju:2020smc,Chen:2019uhq,Hotta:2016qtv}. Researchers continue to explore this fascinating field, uncovering new insights into the nature of black holes and their role in the universe.\\
As our understanding of fundamental physics progressed, it became clear that the classical description of black hole thermodynamics may require modifications to fully encapsulate the underlying quantum gravitational effects and other exotic phenomena \cite{DeWitt:1967yk}. Various theories beyond Einstein's general relativity have been proposed \cite{Clifton:2011jh,DeFelice:2010aj,Berti:2015itd,Cai:2015emx,Palti:2019pca} including complete reviews on modified gravity \cite{Nojiri:2010wj,Nojiri:2017ncd}, encompassing novel fields and interactions that can potentially influence black hole thermodynamics. One such theory is the ENE-dilaton theory \cite{Sheykhi:2014gia}, which extends general relativity by incorporating nonlinear electrodynamics and a dilaton field \cite{NED}. In the presence of a uniform electric field, this theory introduces intriguing modifications to the properties of black holes, including their thermodynamic quantities. Furthermore, black holes in the context of this theory are often referred to as "dirty" due to the presence of additional fields and interactions beyond the vacuum solutions of classical general relativity \cite{Visser:1993nu,Visser:1992qh}.\\
It is widely known that in any thermodynamic system, there are thermal fluctuations occurring at a quantum level. These fluctuations contribute to the system's entropy, along with a logarithmic term that arises from a perturbative correction \cite{Log}. These perturbative corrections are more significant at scales larger than the Planck scale. However, at the Planck scale and smaller, non-perturbative corrections dominate \cite{main-exp}. This means that in thermodynamic systems like black holes, both perturbative and non-perturbative corrections play a role in determining the entropy \cite{Medved:2005vw,Dabholkar:2014ema}. One consequence of this is that an exponential term is added to modify the black hole area entropy through non-perturbative analysis. This correction appears in all quantum theories of gravity. The impact of this correction is negligible when the black hole has a large horizon radius, but becomes significant when the black hole size becomes extremely small.\\
 In a recent article of Mazharimousavi \cite{Mazharimousavi:2023tbn}, the author has successfully addressed key aspects of Einstein's gravity coupled with square-root-a nonlinear electrodynamics and a dilaton field: an ENE-dilaton theory. In that work, field equations have been precisely solved, yielding a unique black hole solution defined by two significant physical parameters: mass and dilaton field parameters. Notably, while the latter represents a constant characterizing the dilaton, the former is an integration parameter. This black hole is non-asymptotically flat (NAF) and exhibits singularity at its center, coinciding with the location of an electric charge. The electric field is radially symmetric and uniform, maintaining a constant electromagnetic invariant. One notable contribution of that study is the determination of the quasi-local conserved mass, denoted as $M_{Q L}$, which was obtained by using the Brown-York formalism \cite{Brown:1992br} since the ADM mass is not applicable to the NAF black holes. Remarkably, in the Schwarzschild limit, this newly defined mass coincides with the ADM mass of the Schwarzschild black hole. The article also explores the thermal stability of the black hole, revealing that it exhibits thermal stability under specific conditions ($0<b^2<1$ or $2<\eta^2$; for more details, we refer the reader to  Ref. \cite{Mazharimousavi:2023tbn}), as evidenced by positive Hawking temperature and heat capacity. This finding underscores the existence of "dirty" black holes surrounded by normal matter fields, challenging the notion of black holes forming in empty space.\\
 Our focus in this article delves into the non-perturbative corrections to the thermodynamics of dirty black holes of the ENE-dilaton theory \cite{Mazharimousavi:2023tbn}. We investigate key thermodynamic quantities of those dirty black holes such as heat capacity, Helmholtz free energy, and internal energy to uncover the effects of the additional fields and interactions introduced by the theory. Additionally, we consider the realm of quantum corrections and examine the modified entropy of these black holes, shedding light on the interplay between quantum effects and classical thermodynamics. So, we provide a comprehensive analysis of the thermodynamic aspects of black holes within the considered theory. Also, we study the intricate connection between quantum effects and black hole entropy, unveiling the quantum-corrected (QC) aspects of thermodynamic behavior. By examining the non-perturbative corrections to thermodynamic quantities, we aim to contribute to a deeper understanding of the nature of dirty black holes in the context of the ENE-dilaton theory. \\
The paper is organized as follows: In Sec. \ref{sec2}, we introduce the dirty black hole spacetime and highlight some of its distinctive features. In Sec. \ref{sec3}, we explore the theoretical framework of thermodynamics in dirty black hole geometry and examine heat capacity, Helmholtz free energy, and internal energy. In Sec. \ref{sec04}, we present the QC-entropy for the dirty black hole within the context of quantum work. Finally, in Sec. \ref{con}, we summarize our results and discuss potential avenues for further research in this intriguing field.

\section{Features of dirty black holes of ENE-dilaton theory}\label{sec2}
In this section, the properties of the metric of the dirty black holes supported by a uniform electric field in the ENE-dilaton theory are studied. The action used in the ENE-dilaton theory is as follows \cite{Mazharimousavi:2023tbn}
\begin{equation}\label{action}
\mathcal{I} = \int {{d^4}x\, \left[ {\mathcal{R}-\frac{1}{2}{\partial _\mu }\psi {\partial ^\mu }\psi  + {e^{ - 2b\psi }}\mathcal{L}(\mathcal{F} )} \right]} ,
\end{equation}
where $ b\ne 0 $ is a free dilaton parameter, $\mathcal{R}$ is the Ricci scalar, $ \mathcal{F} = F_{\mu\nu}F^{\mu\nu} $ is the electromagnetic invariant, $\mathcal{L}(\mathcal{F} ) =\alpha\sqrt { - \mathcal{F}} $ and $ \alpha $ is a dimensionful constant parameter. Varying the action concerning the metric tensor, dilaton scalar field, and gauge potential results in the Einstein field equation, dilaton field equation, and nonlinear electrodynamics-dilaton equation, respectively, \cite{Mazharimousavi:2023tbn}:
\begin{equation}
\mathcal{R_\mu ^\nu } = 2{\partial _\mu }\psi {\partial ^\mu }\psi  + \frac{{\alpha {e^{ - 2b\psi }}}}{{\sqrt { - \mathcal{F}} }}{F_{\mu \lambda }}{F^{\upsilon \lambda }}, \label{2.1}
\end{equation}

\begin{equation}
{\nabla _\mu }{\nabla ^\mu }\psi (r) = \frac{{\alpha b {e^{ - 2b\psi }}}}{2}\sqrt { - \mathcal{F}}, \label{2.2}
\end{equation}

\begin{equation}
d(\frac{{{e^{ - 2b\psi }}}}{{\sqrt { - \mathcal{F}} }}\tilde F) = 0, \label{2.3}
\end{equation}
where $ \tilde F $ is the dual field two-form of $ F $. At this point, a reader may question the significance of the chosen theory and, most importantly, the theory of non-linear electrodynamics considered in this paper can be thought as an ill-defined theory because the Lagrangian in Eq. \eqref{action} contains the square root of $\sqrt{-\mathcal{F}}$. To clarify this issue, let us recall that the pure electric or magnetic Born-Infeld nonlinear electrodynamics \cite{Born:1933pep} is described by
\begin{equation}
L=b^{2}\left(1-\sqrt{1+\frac{\mathcal{F}}{2b^{2}}}\right), \label{2.5}
\end{equation}
in which $\mathcal{F}=F_{\mu\nu}F^{\mu\nu}$. While the weak field limit of
the theory ($\frac{\mathcal{F}}{b^{2}}\rightarrow0$) is the linear Maxwell
electrodynamics i.e., 
\begin{equation}
L\rightarrow-\frac{1}{4}F_{\mu\nu}F^{\mu\nu}, \label{2.6}
\end{equation}
and its strong field limit $\left(\frac{\mathcal{F}}{b^{2}}\rightarrow large\right)$ yields
\begin{equation}
L\sim\sqrt{\mathcal{F}}, \label{2.7}
\end{equation}
up to a constant coefficient. Therefore, the square-root model is
not supposed to be considered instead of the linear electrodynamics
theory but it is a model for the strong fields. Its applications can
be seen in the literature from the pioneering works of G.~'t Hooft \cite{tHooft:2002pmx}
 and H.~B.~Nielsen and P.~Olesen
\cite{Nielsen:1973qs}. In the sequel, important studies for the applications of the
same model have been published by E. Guendelman and his colleagues \cite{Gaete:2006xd,Gaete:2007ry,Guendelman:2011sm,Guendelman:2012sv,Guendelman:2018zcb,Guendelman:2013sca}, mainly on the confinement of quarks,  and their followers like S.~H.~Mazharimousavi, M.~Halilsoy, and A.~\"Ovg\"un \cite{Mazharimousavi:2012zx,Ovgun:2021ttv,Mazharimousavi:2022lji}
 and references therein. Therefore, this ENE-dilaton model has a solid foundation and whence, in such a configuration, either there is a strong magnetic
field or a strong electric field, one can set $L\sim\sqrt{\pm \mathcal{F}}$ \cite{Gaete:2006xd,Gaete:2007ry,Guendelman:2011sm,Guendelman:2012sv,Guendelman:2018zcb,Guendelman:2013sca,Mazharimousavi:2012zx,Ovgun:2021ttv,Mazharimousavi:2022lji}.

Based on Eqs. \eqref{2.1}-\eqref{2.3}, the metric for a static, spherically symmetric black hole spacetime was expressed as \cite{Mazharimousavi:2023tbn}:
\begin{equation}\label{st1}
d{s^2} =  - {\eta ^2}\left( {1 - {{\left( {\frac{{{r_ + }}}{r}} \right)}^{{\eta ^2}}}} \right){r^{\frac{2}{{{b^2}}}}}d{t^2} + \frac{{{\eta ^2}}}{{1 - {{\left( {\frac{{{r_ + }}}{r}} \right)}^{{\eta ^2}}}}}d{r^2} + {r^2}(d{\theta ^2} + {\sin ^2}\theta d{\varphi ^2})	,
\end{equation}
where $r_{+}$ represents the event horizon of a black hole described by the above spacetime and $\eta^{2}=\frac{b^{2}+1}{b^{2}}>1$ is the dilaton parameter. Furthermore, Eq. (\ref{st1}) can be re-expressed as follows:
\begin{equation}\label{metric}
 d{s^2} =  - (1-\frac{{2M}}{{{\rho ^{{\eta ^2}}}}}){\rho ^{2({\eta ^2} - 1)}}d{\tau^2} + (1 - \frac{{2M}}{{{\rho ^{{\eta ^2}}}}})^{ - 1}d{\rho^2} + \dfrac{\rho^{2}}{\eta^{2}}(d{\theta^{2}}+{\sin ^2}\theta d{\varphi ^2}) , 
\end{equation}
where $ \rho = \eta r $, $ M = \frac{{{{(\eta {r_ + })}^{{\eta ^2}}}}}{2} $, and $ \tau  = {\eta ^{1 - \frac{2}{{{b^2}}}}}t $.

The Hawking temperature of the dirty black hole \eqref{metric} can be computed with the aid of a timelike Killing vector ($\chi^{\mu}$) and whence the surface gravity ($\kappa$):
\begin{equation}
T_H=\frac{\kappa}{2 \pi}=\left. \frac{\nabla_\mu \chi^\mu \nabla_\nu \chi^\nu}{2 \pi}\right|_{r=r_{+}}=\frac{\eta^2}{4 \pi} r_{+}^{\eta^2-2}, \label{hawt}
\end{equation}
and when the black hole area law \cite{5-0-B} is applied, the black hole's entropy can be determined as
\begin{equation}
 S_{BH}=\pi r_{+}^2. \label{sbh}  
\end{equation}
It is also worth noting that the NAF structure of metric \eqref{st1} admits the quasilocal mass \cite{Brown:1992br} as $M_{QL}=\frac{r_{+}^{\eta^2}}{2}$ in which $\eta^2$ is related to the background. Thus, the first law of thermodynamics of the dirty black hole is satisfied as:
\begin{equation}
dM_{Q L}=T_H dS_{BH}.    
\end{equation}

\section{Thermodynamics}\label{sec3}
In this section, we will study the thermodynamics of a dirty black hole supported by a uniform electric field in the ENE-dilaton theory by applying quantum corrections to the entropy.

\subsection{Exponential correction}\label{exp}
Researchers have used two types of quantum corrections: perturbative (\cite{BM,BSSM,BSHH,BMZA,BAI,Upadhyay:2018bqy}) and non-perturbative (\cite{SYR,AA,Soroushfar:2023mrm,Pourhassan:2022sfk,Pourhassan:2021mhb}). These corrections have been applied to analyze and investigate the thermodynamic properties of various black holes. It is well-known that the entropy of a system (denoted by $S_{0}$) is related to the number of measurable microstates ($\Omega$). This relation can be expressed in units of Boltzmann constant ($k_{B}=1$) as follows:
\begin{equation}\label{SO}
S_{0}\equiv S_{BH}=\ln{\Omega}.
\end{equation}
However, there are other unmeasurable microstates that contribute to the entropy ($S_{micro}$). The probability of finding such states is inversely proportional to $\Omega$, and the corresponding entropy is proportional to the probability. Hence,
\begin{equation}\label{Sm}
S_{micro}\propto\frac{1}{\Omega},
\end{equation}
Therefore, one can write,
\begin{equation}\label{St}
S=S_{0}+S_{micro}.
\end{equation}
Combining Eqs. (\ref{SO}) and (\ref{Sm}), we get
\begin{equation}\label{SmO}
S_{micro}=e^{-S_{0}}.
\end{equation}
Hence, using Stirling's approximation and statistical physics for a large total number $N$, the exponential quantum correction to black hole entropy is given by:
\begin{equation}\label{S}
S=S_{0}+e^{-S_{0}},
\end{equation}
which is consistent with non-perturbative aspects of string theory \cite{string,Nielsen:1973qs} and applicable to the Planckian regime of black hole event horizon area. Next, we apply the non-perturbatively corrected entropy to check the thermodynamic properties of the black hole resulting from these changes in entropy \cite{Deh, Beh}:
\begin{equation}
S = S_0  + \lambda {e^{ - S_0}}, \label{non}
\end{equation}
where $\lambda$ is the correction coefficient which is related to the proportionality constant in Eq. (\ref{Sm}). So, we have
\begin{equation}\label{SC}
S = \pi  r_{+}^{2} + \lambda {e^{ - \pi  r_{+}^{2}}},
\end{equation}
which means that the last term of Eq. (\ref{SC}) affects the black hole thermodynamic quantities.

\subsection{Heat capacity}\label{heat}
One can compute the standard specific heat capacity of the dirty black hole as follows (see Eqs. \eqref{hawt} and \eqref{sbh}):
\begin{equation}
 C_{BH}=T_H \frac{\partial S_{BH}}{\partial T_H}=\frac{2 \pi r_{+}^2}{\eta^2-2}.   
\end{equation}
If both $T_H$ and $C_{BH}$ are positive, the black hole is considered thermally stable. Hence, when $\eta^2-2>0\left(b^2<1\right)$, the black hole is thermally stable. When $\eta^2=2$, the Hawking temperature \eqref{hawt} remains constant, which corresponds to the infinite heat capacity. Similarly, one can compute the corrected heat capacity $\left( C=T_{H} \frac{{\partial S}}{{\partial T_{H}}}\right)$ by employing Eqs. \eqref{hawt} and \eqref{SC} for the dirty black hole as follows
\begin{equation}\label{C}
C=\frac{2\pi r_{+}^{2} \left(\lambda  \,{\mathrm e}^{-\pi  r_{+}^{2}}-1\right)}{2-\eta^{2}},
\end{equation}
 which reduces to the original heat capacity $C_{BH}$ in the case of $ \lambda  = 0 $.\\
\begin{figure}[h]
	\centering
	\subfigure[$ \eta=2.2 $]{
		\includegraphics[width=0.450\textwidth]{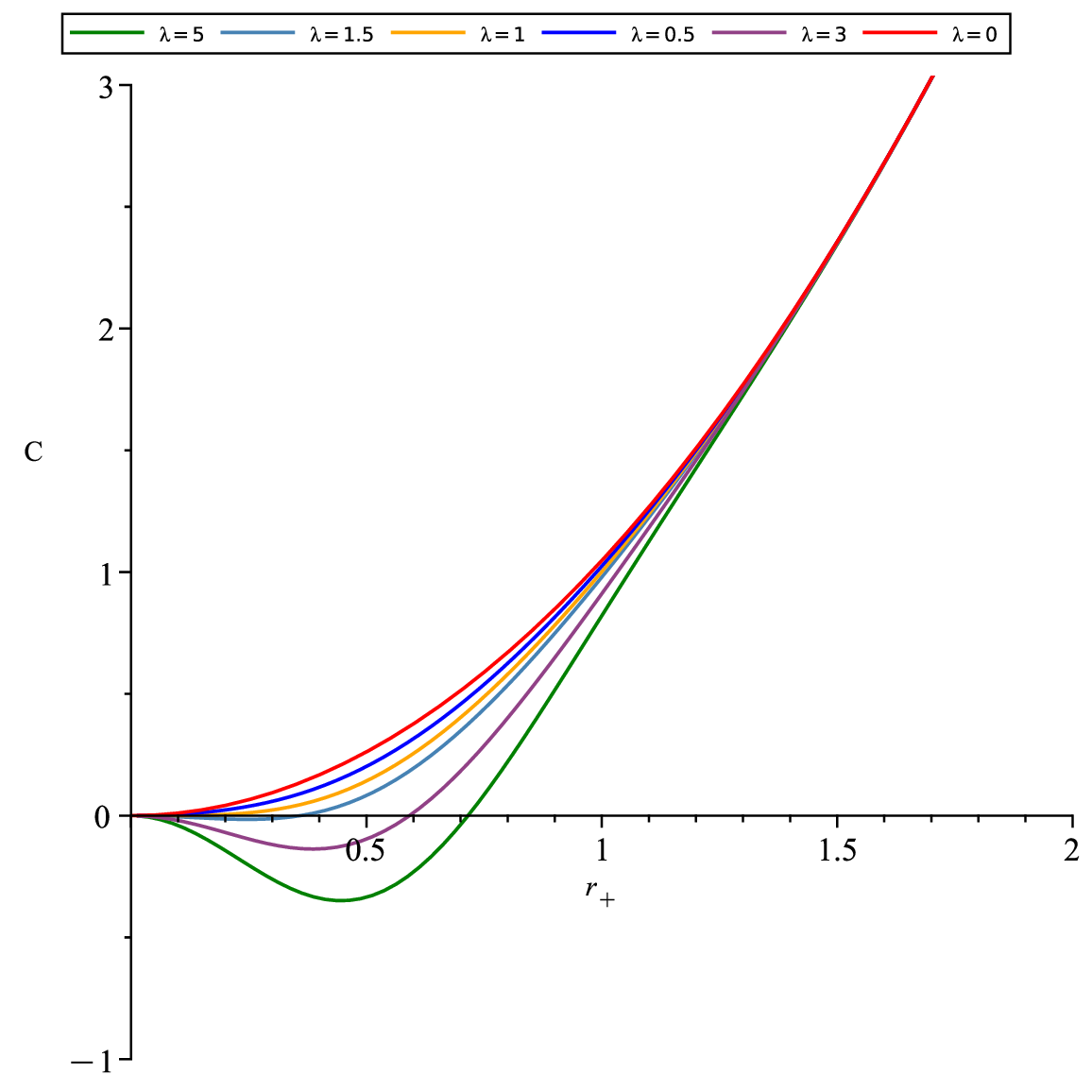}
	}
\subfigure[closeup of (a) ]{
	\includegraphics[width=0.450\textwidth]{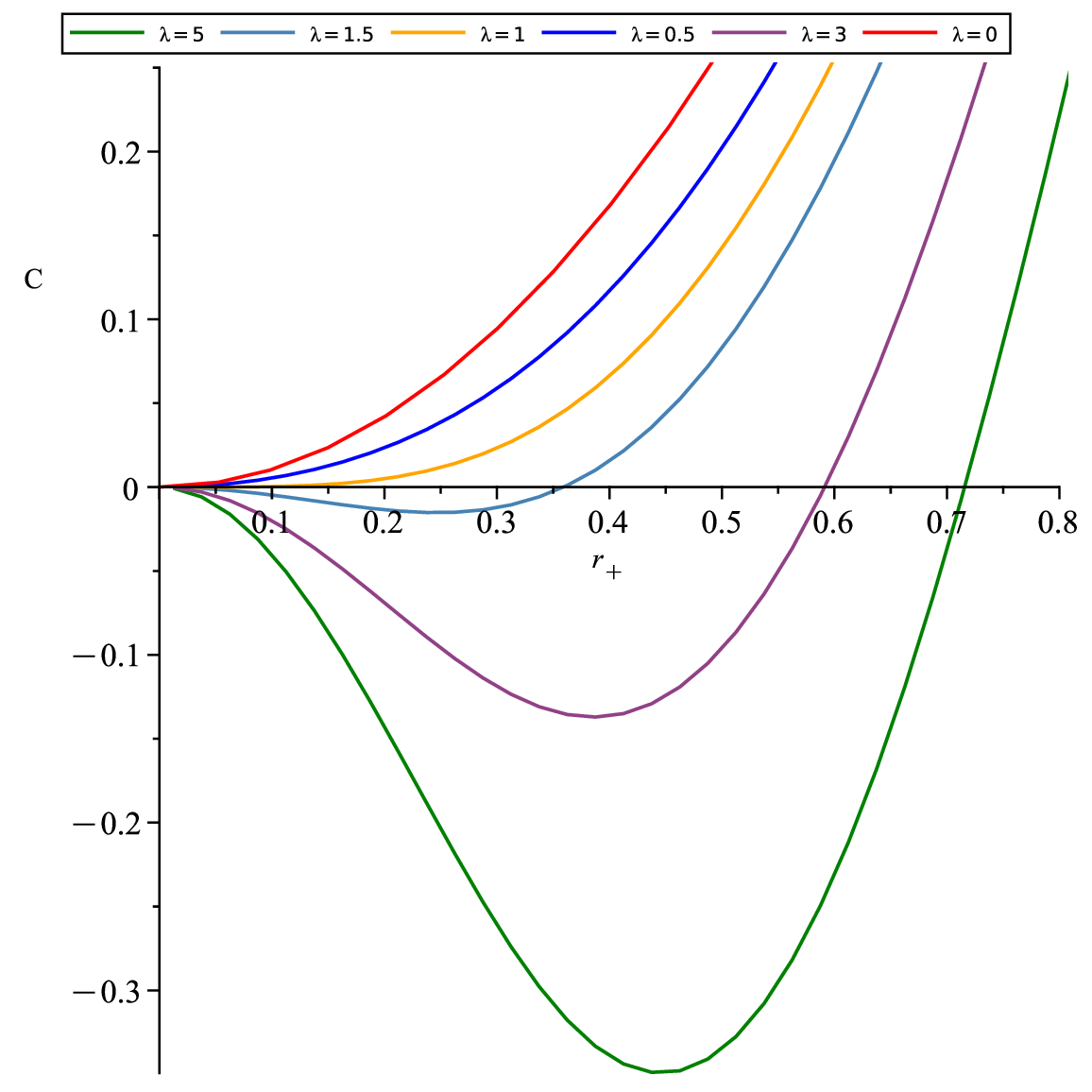}
}
\subfigure[$ \lambda=5 $]{
	\includegraphics[width=0.450\textwidth]{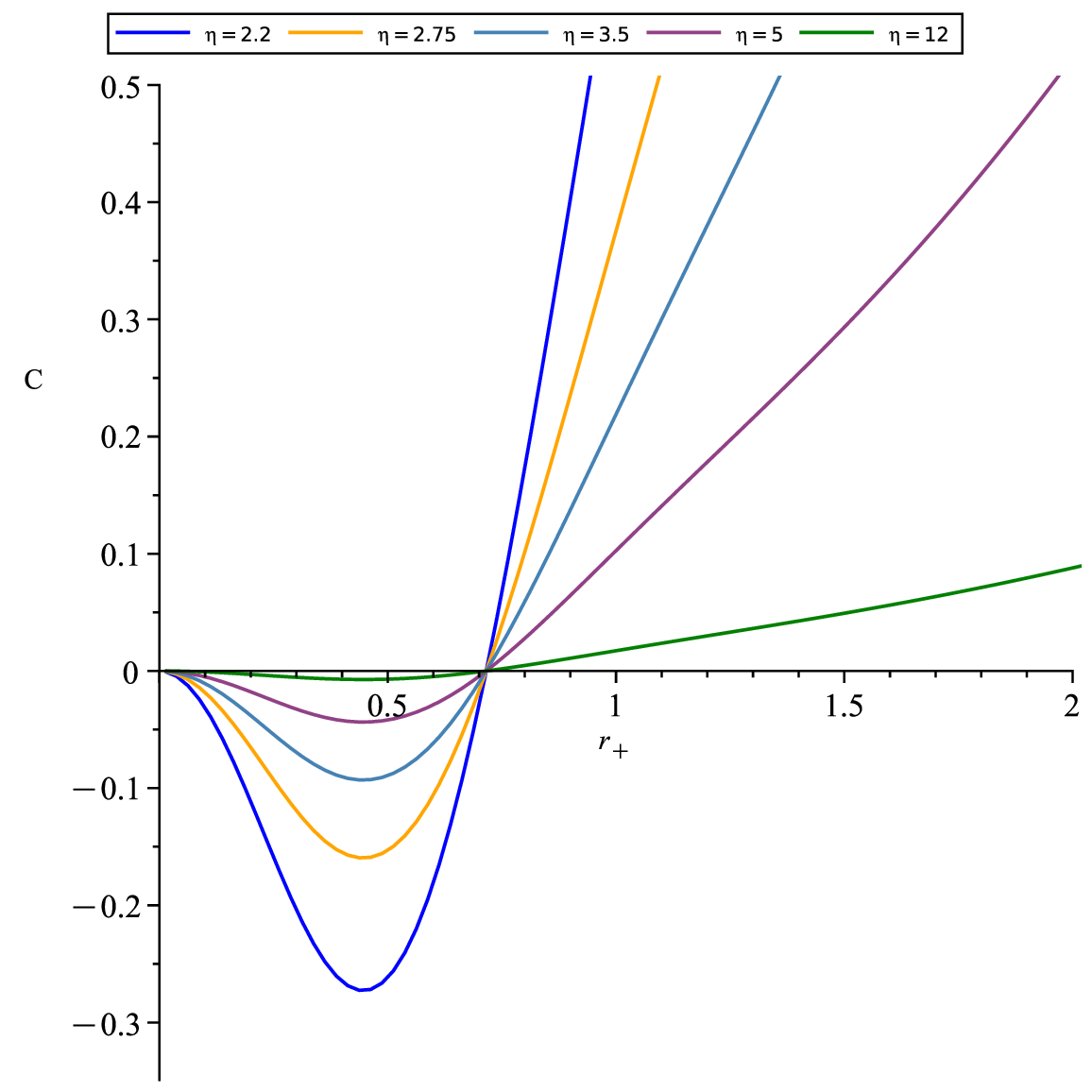}
}
\subfigure[$ \lambda=-5 $]{
	\includegraphics[width=0.450\textwidth]{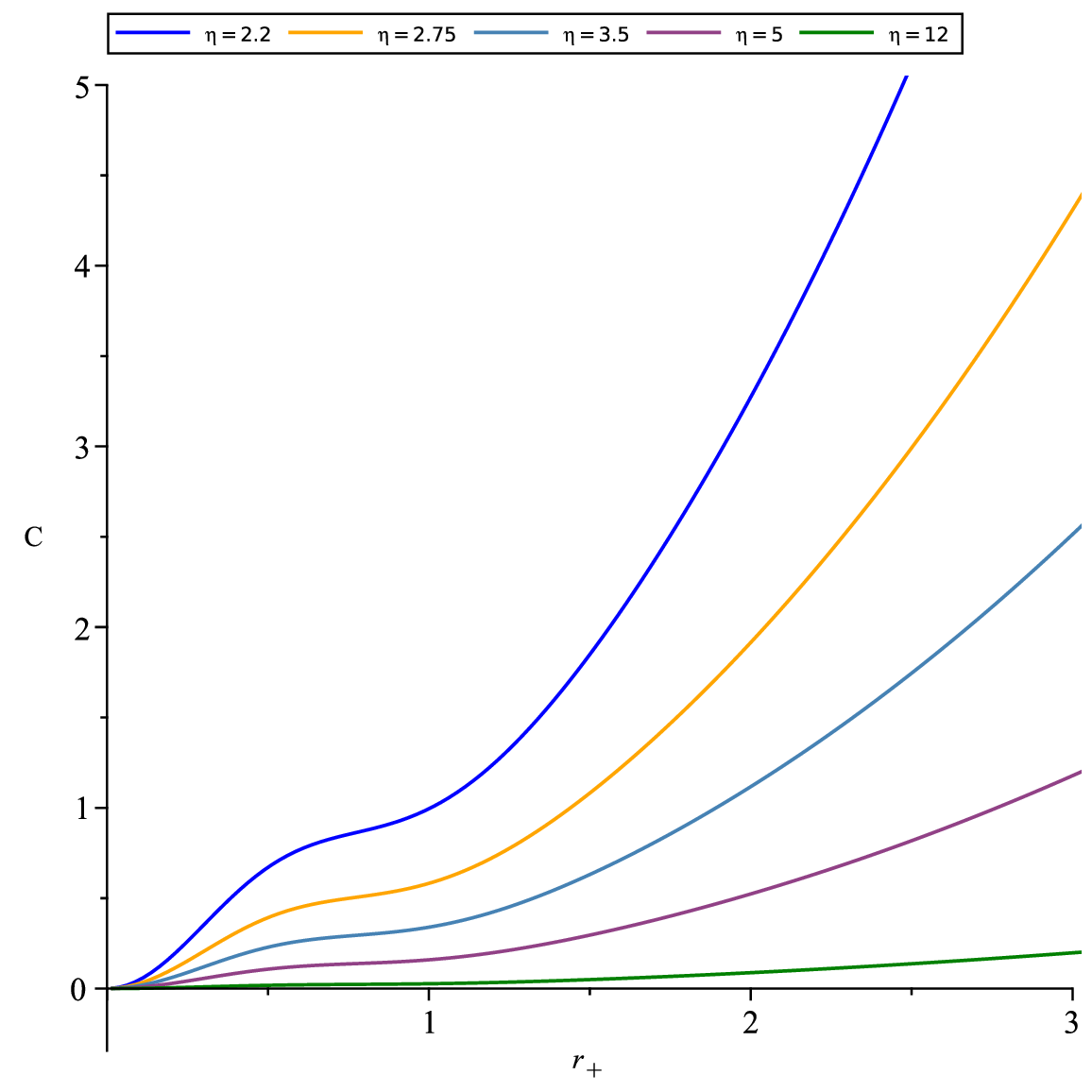}
}
	\caption{Variations of the original and the corrected heat capacity according to the horizon radius $ r_{+} $.}
	\label{Pic:C}
\end{figure}

Figs. \ref{Pic:C} (a)-(d) illustrate the behavior of a dirty black hole's heat capacity when supported by a uniform electric field according to the ENE-dilaton theory for various values of $ \lambda $ and $ \eta $. It can be observed from Figs. \ref{Pic:C} (a) and (b) that the original heat capacity ($\lambda =0 $) of this black hole is positive and has no phase transition. For the QC heat capacity, ($\lambda \ne 0$), and by increasing the values of $\lambda$, the heat capacity enters the negative region (unstable phase), then takes phase transition type one, and afterward it will be positive (stable) again. Hence, by reducing the black hole size due to Hawking radiation, the final stage of this black hole leads to instability (for the positive correction parameter).

Moreover, the behavior of the QC heat capacity for different values of the dilaton parameter ($\eta $) is shown in Figs. \ref{Pic:C} (c) and (d). In Fig. \ref{Pic:C} (c), we find that, for a certain positive value of $\lambda$, the system has phase transition type one, but for a certain negative value of $\lambda$, the system is in the stable phase and it has no phase transition [see Fig. \ref{Pic:C} (d)]. In this case, the final stage of the black hole is stable as well as uncorrected. In addition, from Fig. \ref{Pic:C} (c), it is clear that, by increasing the values of $\eta$, the QC heat capacity gradually tends to zero.

In any case, by looking at the aforementioned diagrams [Figs. \ref{Pic:C} (a)-(d)], analyzing them, and examining the variations in heat capacity for various values of $\lambda$ and $\eta$, as well as taking into account the phase transition that takes place for the heat capacity in the quantum correction mode, it can be deduced that a small black hole could be unstable due to quantum effects (assuming positive correction parameter), but a large black hole is in a thermodynamically stable phase. In the following subsections, we shall look at how the aforementioned quantum corrections affect the internal energy and Helmholtz free energy, among other thermodynamic variables.

\subsection{Helmholtz free energy}\label{Helmholtz}
In simple terms, Helmholtz free energy is a thermodynamic potential that combines internal energy and entropy, providing insight into a system's ability to perform work at constant temperature and volume \cite{Sadeghi:2016dvc}. When applied to black holes, it offers a unique perspective on their behavior. The Helmholtz free energy is used to understand the equilibrium between a black hole and its surrounding radiation, shedding light on the intricate interplay between mass, temperature, and entropy in these mysterious cosmic objects.

\begin{figure}[h]
	\centering
	\subfigure[$\eta =2.5 $]{
		\includegraphics[width=0.450\textwidth]{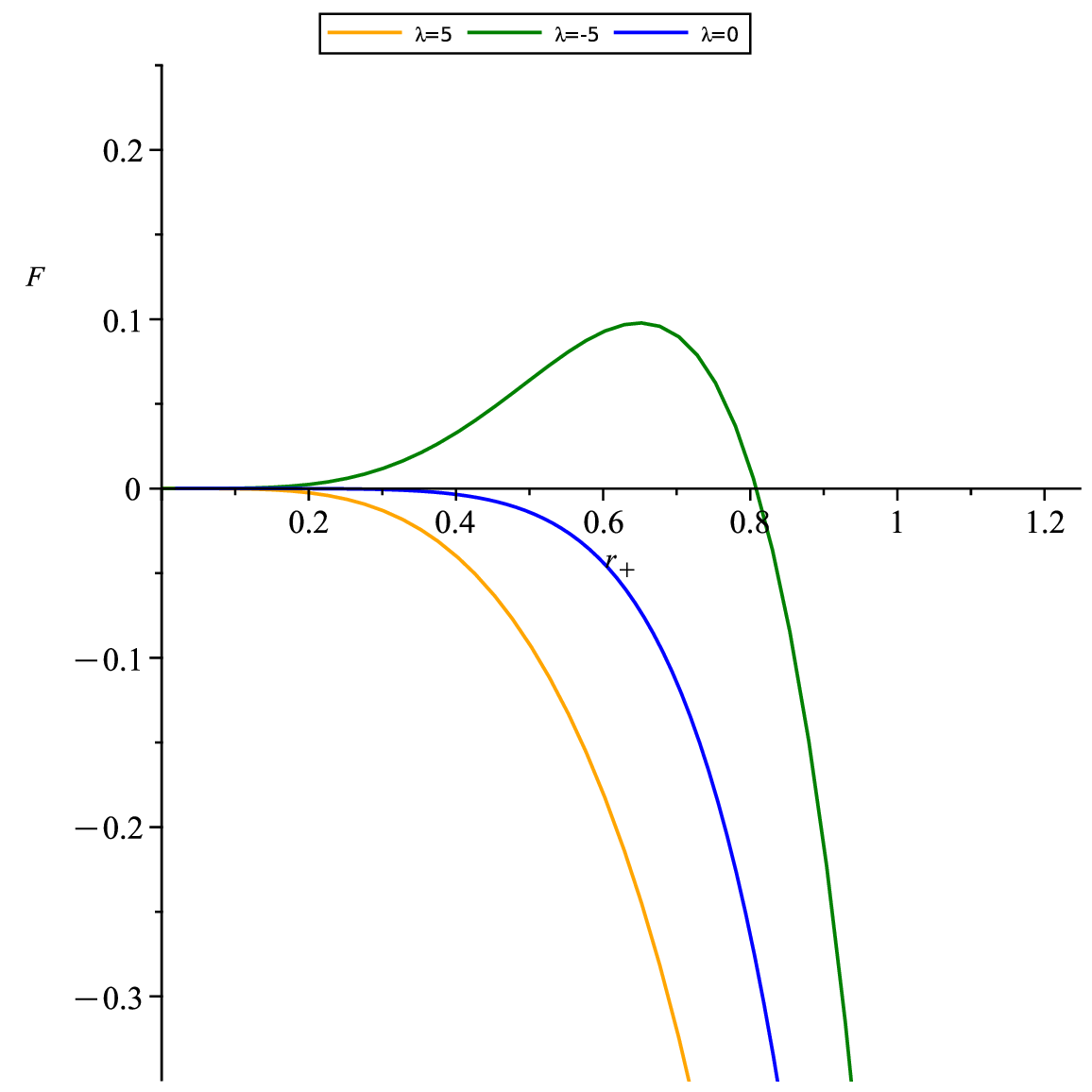}
}
\subfigure[$ \lambda=5 $]{
	\includegraphics[width=0.450\textwidth]{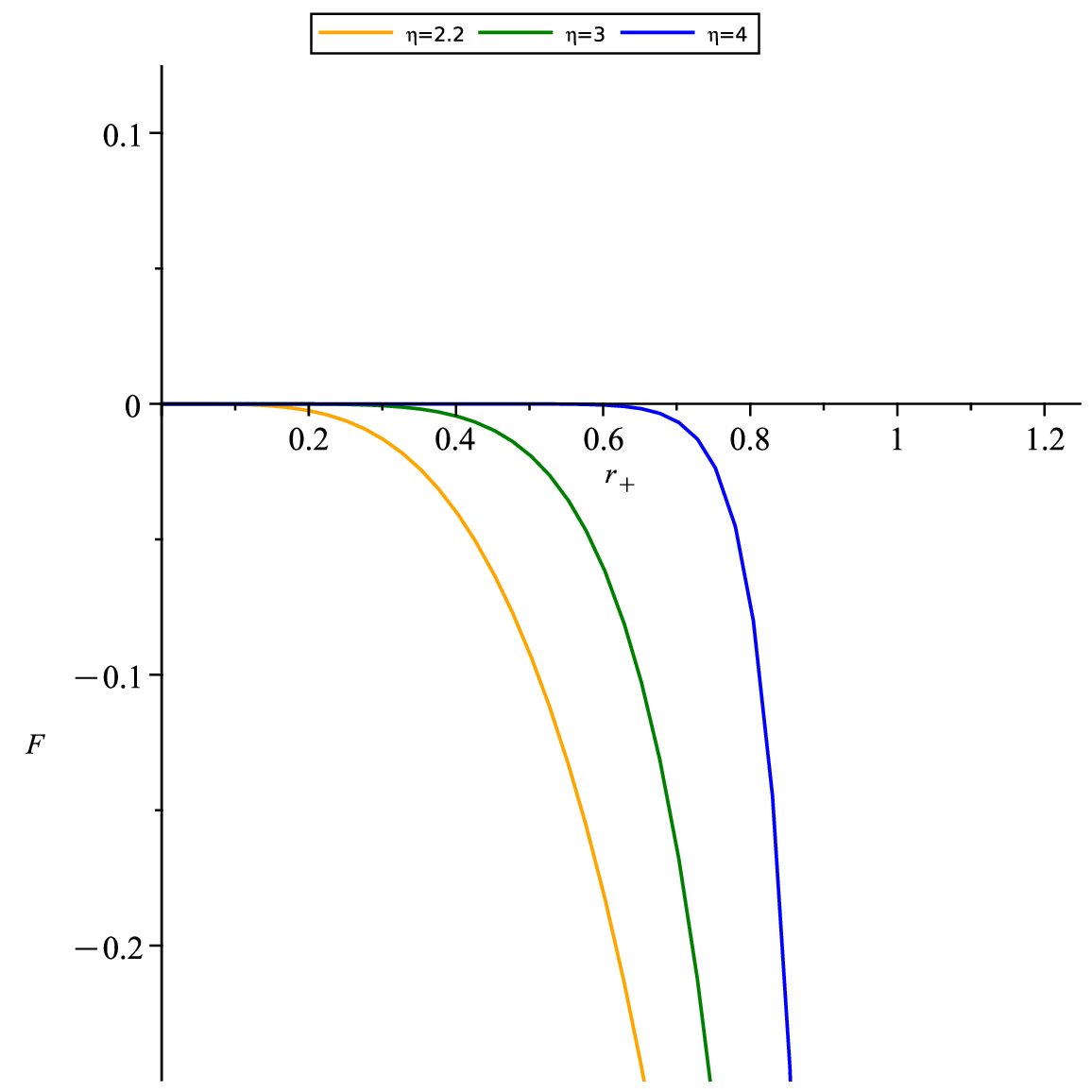}
}
\subfigure[$ \lambda=-5 $]{
	\includegraphics[width=0.450\textwidth]{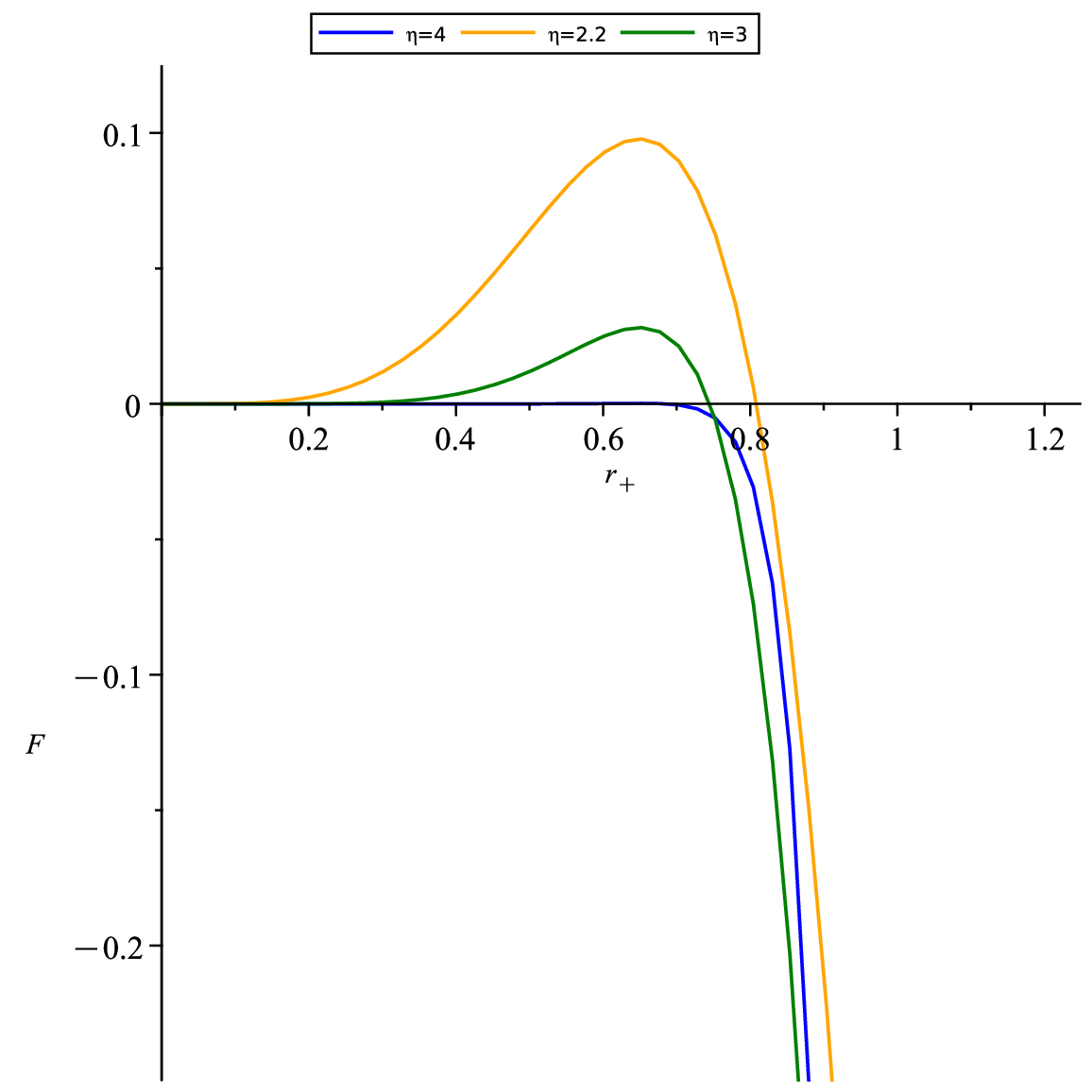}
}
	\caption{Variations of the Helmholtz free energy in terms of horizon radius $ r_{+} $.}
	\label{Pic:F}
\end{figure}

Here, the effects of quantum correction on the Helmholtz free energy, which can be fruitful in analyzing the stability and phase transition of a black hole, are investigated. The Helmholtz free energy is given by \cite{Dolan:2010ha}:
\begin{equation}\label{FA}
F =  - \int {SdT}.
\end{equation}
So, by employing Eqs. (\ref{SO}), (\ref{SC}), and (\ref{FA}), one can get
\begin{eqnarray}\label{F}
F=&-&\frac{r_{+}^{\eta^{2}} \pi^{1-\frac{\eta^{2}}{4}} {\mathrm e}^{-\frac{\pi  r_{+}^{2}}{2}} M_{W} \! \left({\frac{\eta^{2}}{4},\frac{\eta^{2}}{4}+\frac{1}{2}}, \pi  r_{+}^{2}\right) \left(r_{+}^{2}\right)^{-\frac{\eta^{2}}{4}} \lambda}{\pi  \left(\eta^{2}+2\right)}\nonumber\\
&-&\frac{\left(\lambda  \,\eta^{2} r_{+}^{\eta^{2}-2}+2 \lambda  r_{+}^{\eta^{2}} \pi \right) {\mathrm e}^{-\pi  r_{+}^{2}}+r_{+}^{\eta^{2}} \pi  \left(\eta^{2}-2\right)}{4 \pi},
\end{eqnarray}
in which $M_{W}$ denotes the $ WhittakerM(\mu,\nu,z) $ function and it can be defined in terms of the hypergeometric function \cite{ASbook,Lukebook}. In addition, the original Helmholtz free energy of the dirty black hole, in the case of $ \lambda=0 $, can be written as follows
\begin{equation}\label{F0}
F_{0}= -\frac{r_{+}^{\eta^{2}} \left(\eta^{2}-2\right)}{4}.
\end{equation}
The behaviors of the Helmholtz free energy for different values of $ \lambda $ and $ \eta $ are plotted in Fig. \ref{Pic:F}.

Figure \ref{Pic:F} (a) illustrates that, for a specific positive $\eta$ and $\lambda \ge 0$, the Helmholtz free energy is consistently negative, whereas it becomes positive for $\lambda < 0$. Additionally, in Fig. \ref{Pic:F} (b), we observe that, with a certain positive $\lambda$, the Helmholtz free energy remains negative across all values of $\eta$. Conversely, in Fig. \ref{Pic:F} (c), a specific negative $\lambda$ yields a positive Helmholtz free energy value, which subsequently decreases as $\eta$ values increase.

\subsection{Internal energy}\label{internal}
One of the most intriguing aspects of black holes is their thermodynamic behavior, which is described by analogies to classical thermodynamics. The concept of internal energy becomes essential in this context. In the study of black holes, internal energy refers to the energy contained within the black hole itself, often associated with the mass-energy equivalence principle, $E=Mc^2$. The process of Hawking radiation causes the black hole to lose mass over time and, consequently, its internal energy.

\begin{figure}[h]
	\centering
	\subfigure[$\eta =2.5 $]{
		\includegraphics[width=0.450\textwidth]{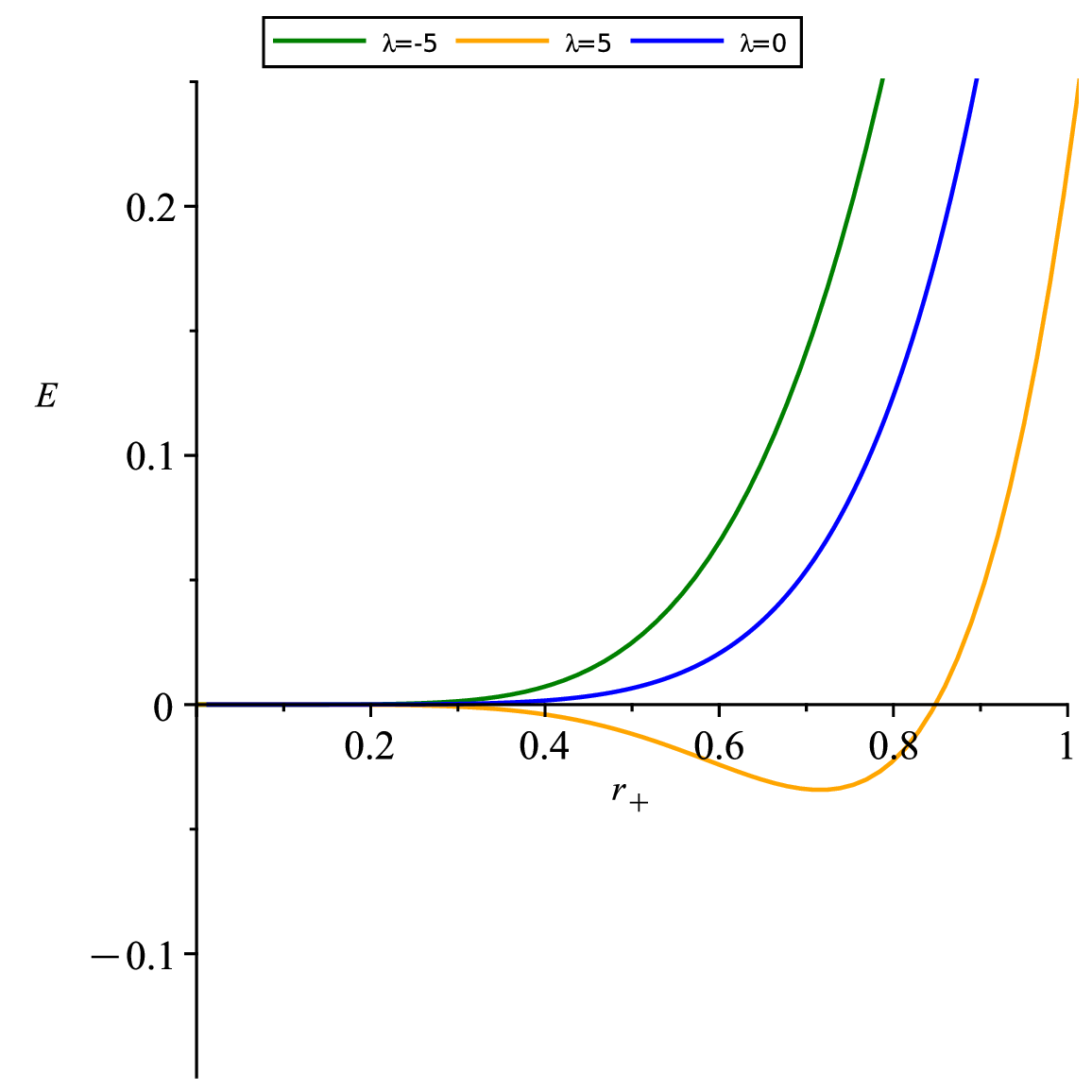}
	}
	\subfigure[$ \lambda=5 $]{
		\includegraphics[width=0.450\textwidth]{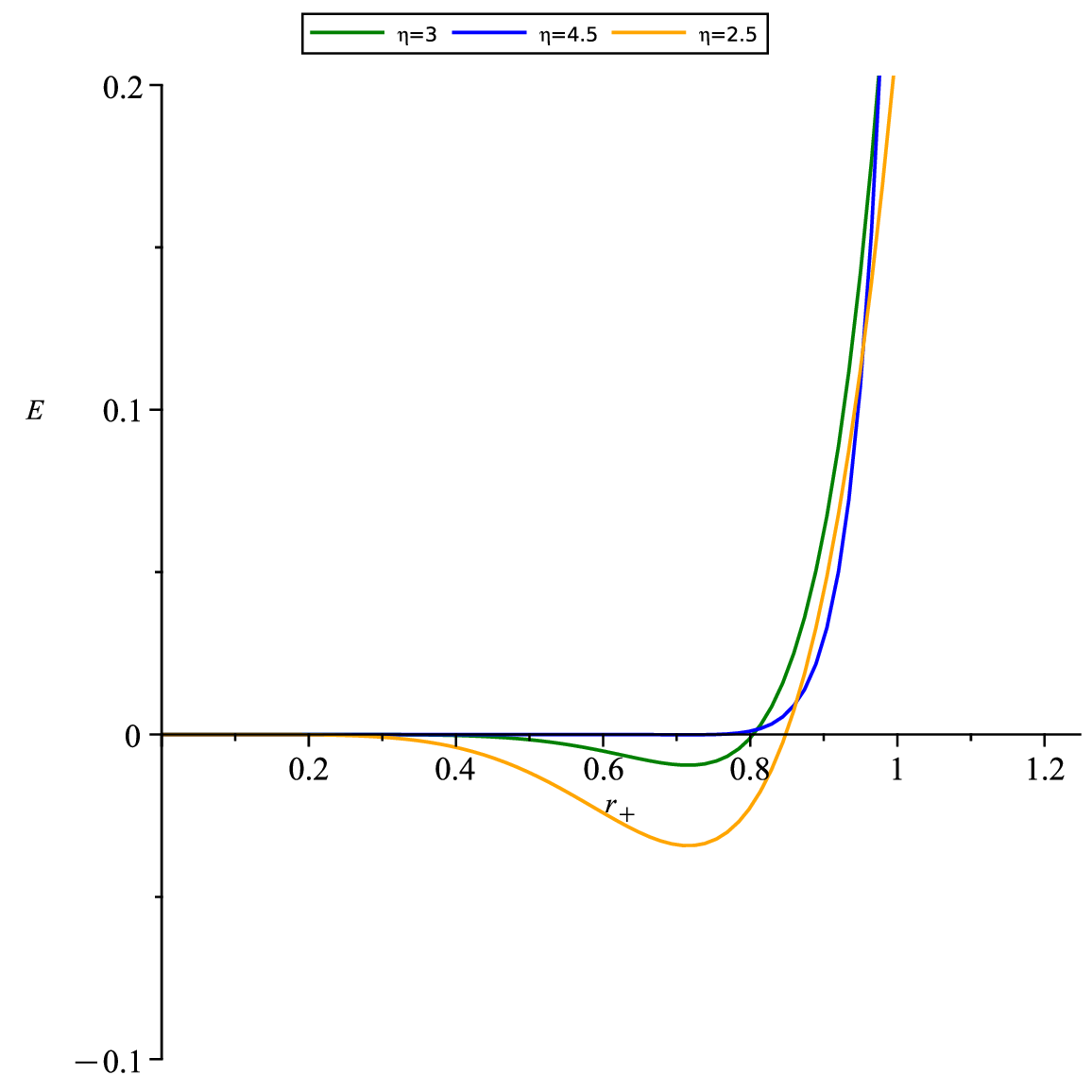}
	}
	\subfigure[$ \lambda=-5 $]{
		\includegraphics[width=0.450\textwidth]{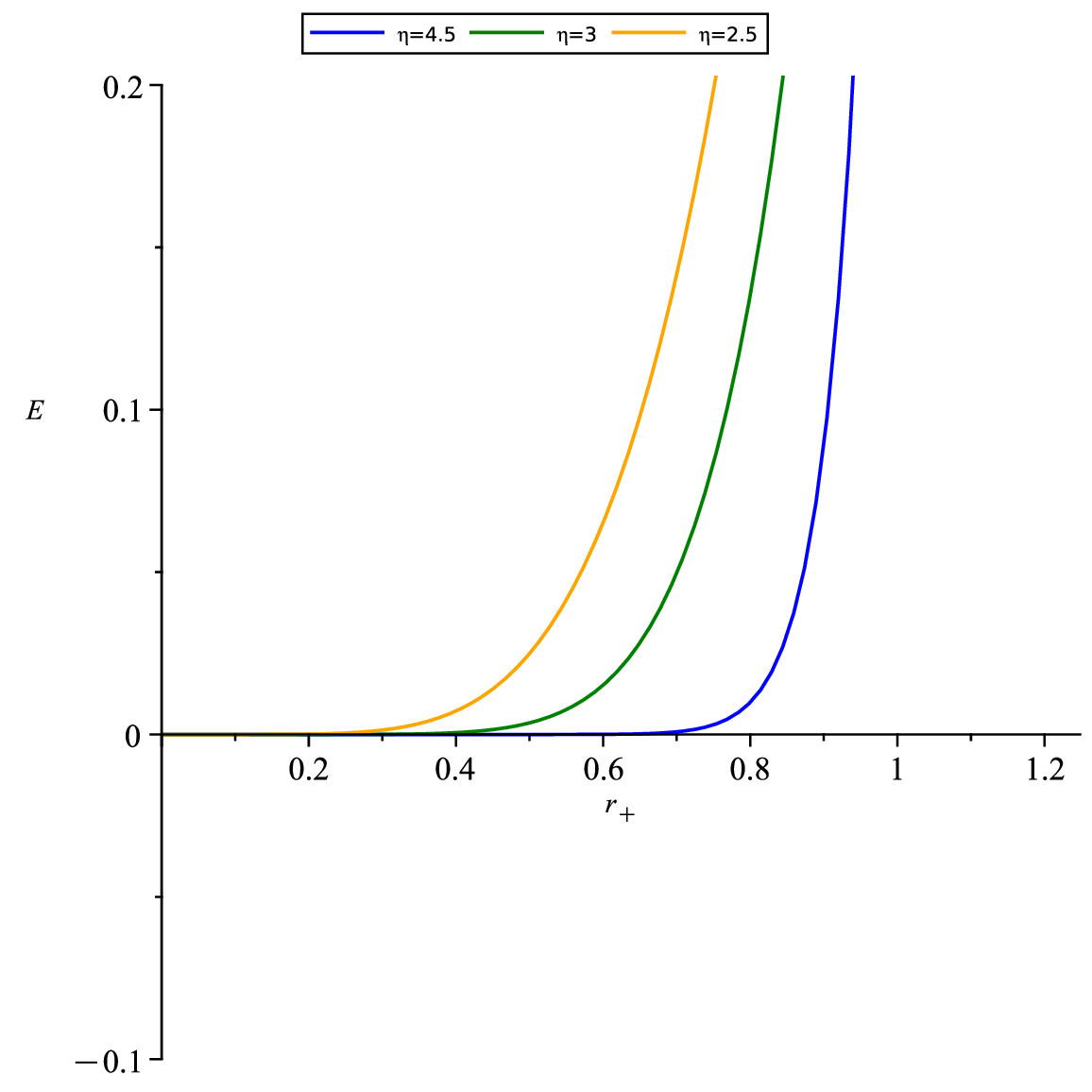}
	}
	\caption{Variations of the internal energy in terms of horizon radius $ r_{+} $.}
	\label{Pic:E}
\end{figure}

In this subsection, the effects of quantum correction on the internal energy is investigated.
The general expression for the internal energy of a black hole is given by \cite{Hayward:1997jp}
\begin{equation}\label{E0}
E = \int {T_{H}dS}.
\end{equation}
Therefore, by using the Eqs. (\ref{SO}), (\ref{SC}) and (\ref{E0}), we have
\begin{equation}\label{E}
E = -\frac{2 \left({\mathrm e}^{-\frac{\pi  r_{+}^{2}}{2}} M_{W} \! \left({\frac{\eta^{2}}{4},\frac{\eta^{2}}{4}+\frac{1}{2}}, \pi  r_{+}^{2}\right) \pi^{-\frac{\eta^{2}}{4}}\left(r_{+}^{2}\right)^{-\frac{\eta^{2}}{4}} \lambda +\frac{\left(\eta^{2}+2\right) \left(\lambda  \,{\mathrm e}^{-\pi  r_{+}^{2}}-1\right)}{2}\right) r_{+}^{\eta^{2}}}{2 \eta^{2}+4}
.
\end{equation}
Therefore, the standard (when $ \lambda=0 $) internal energy can be obtained as
\begin{equation}
E_{0}=\frac{r_{+}^{\eta^{2}}}{2}.
\end{equation}
The plots of the internal energy for different values of $ \lambda $ and $ \eta $, according to the horizon radius are depicted in Fig. \ref{Pic:E}.

Referring to Fig. \ref{Pic:E} (a), it becomes evident that, for a specific positive value of $\eta$ and when $ \lambda \le 0 $, the internal energy consistently remains positive. However, when $ \lambda > 0 $, the internal energy takes on negative values. Furthermore, Fig. \ref{Pic:E} (b) reveals that, with a certain positive value of $\lambda$, the internal energy turns negative, and as the parameter $\eta$ increases, the internal energy gradually converges to zero. In addition, Fig. \ref{Pic:E} (c) illustrates that, for a particular negative value of $\lambda$, the internal energy remains positive across all values of $\eta$.
 
\section{Quantum work}\label{sec04}
The quantum-corrected entropy change of a dirty black hole subjected to a uniform electric field in the ENE-dilaton theory is expressed as \cite{Pourhassan:2022opb}:

\begin{equation}\label{Delta S}
\Delta S = S_{f} - S_{i},
\end{equation}

where $S_{i}$ represents the initial entropy during the evolution and $S_{f}$ stands for the final entropy. Therefore, one can find out

\begin{equation}\label{Delta S2}
\Delta S = \lambda \left( ,{\mathrm e}^{-\pi r_{+f}^{2}} - ,{\mathrm e}^{-\pi r_{+i}^{2}}\right) + \pi \left(r_{+f}^{2} - r_{+i}^{2}\right).
\end{equation}

\begin{figure}[h]
\centering
\subfigure[$\eta = 2.5 $]{
\includegraphics[width=0.450\textwidth]{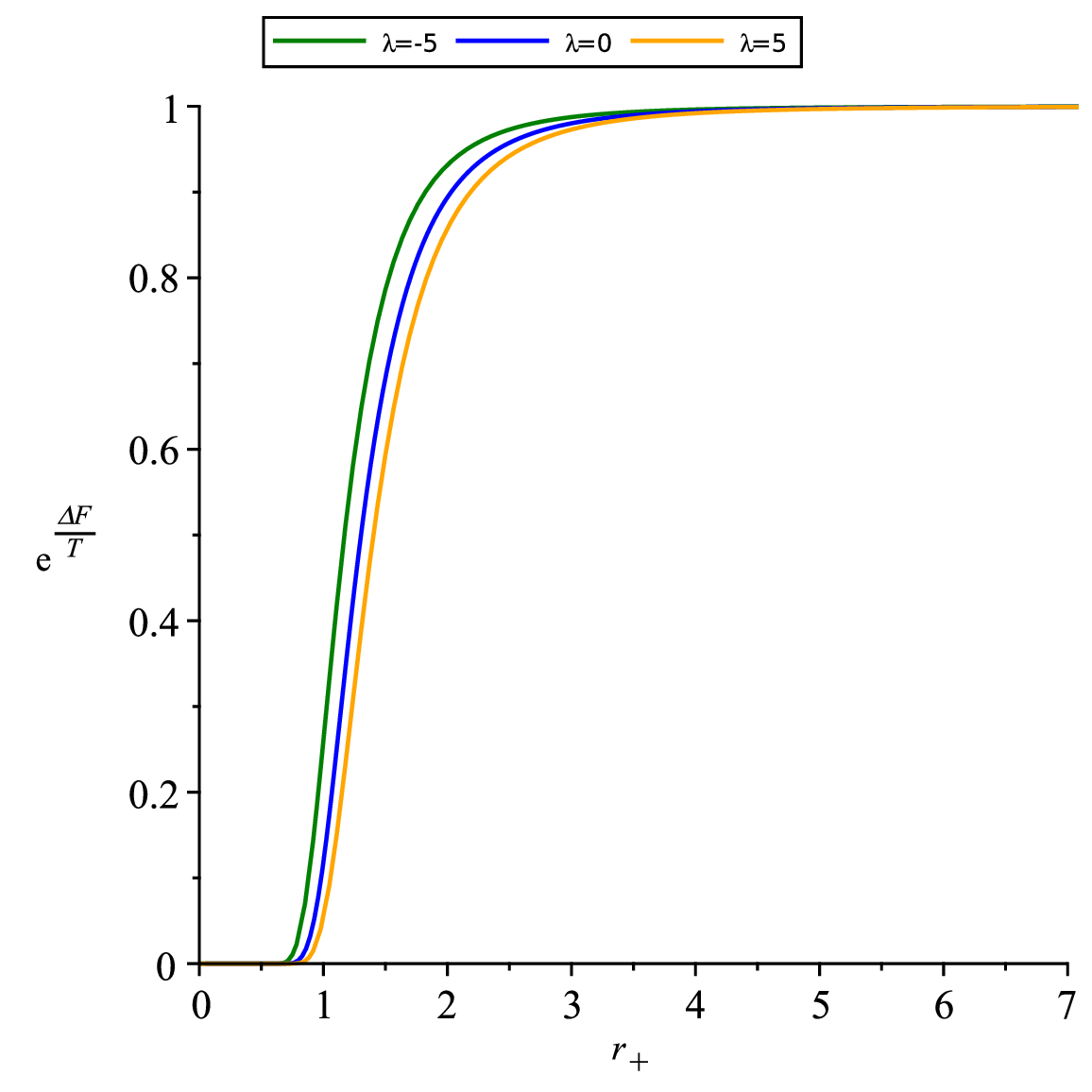}
}
\subfigure[$ \lambda = 5 $]{
\includegraphics[width=0.450\textwidth]{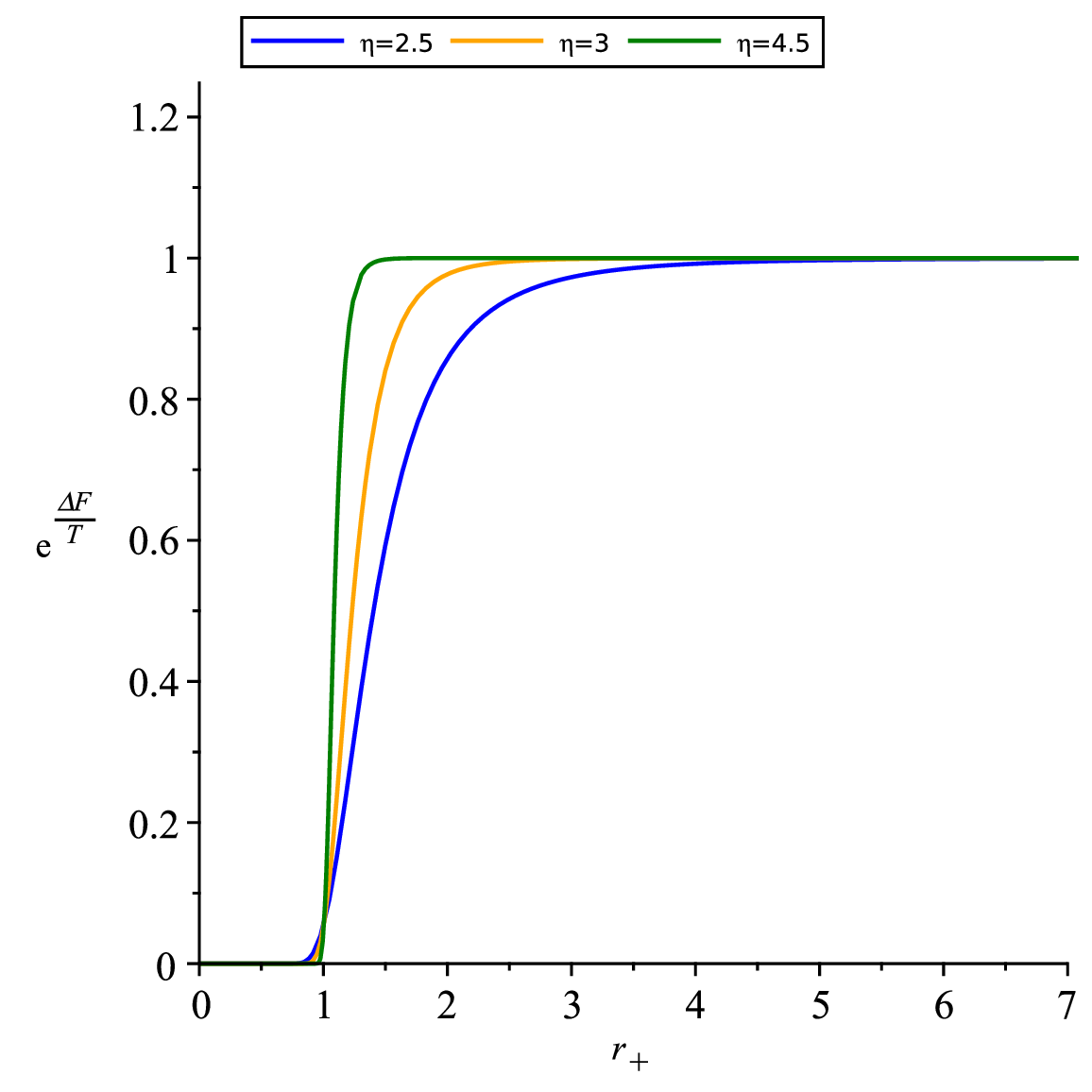}
}
\subfigure[$ \lambda = -5 $]{
\includegraphics[width=0.450\textwidth]{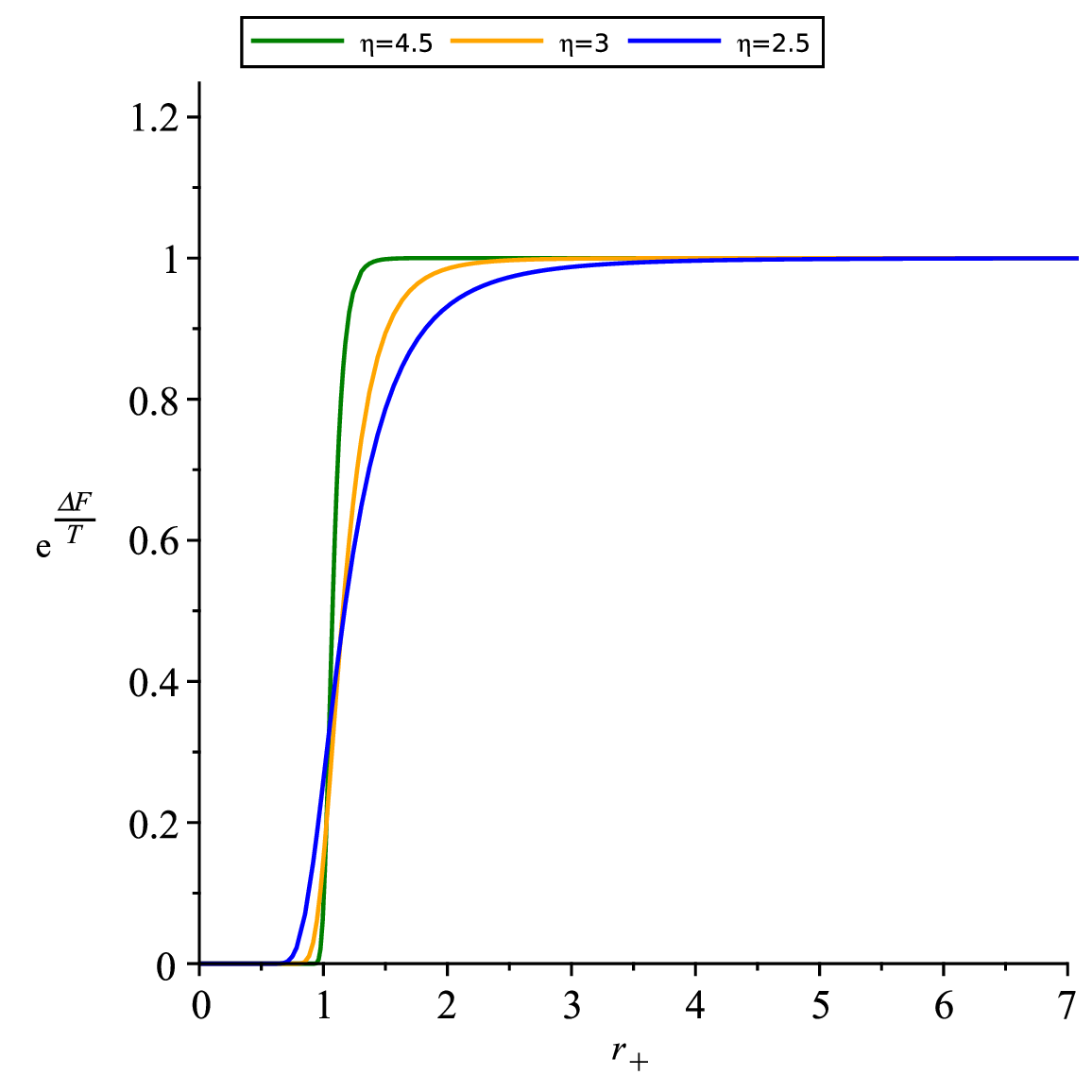}
}
\caption{Variations of quantum work ($e^{\frac{\Delta F}{T}}$) with respect to the horizon radius $ r_{+} $.}
\label{Pic:W}
\end{figure}

Furthermore, to analyze the quantum work involved, it is beneficial to consider the change in the Helmholtz free energy. Thus, we can express the change in Helmholtz free energy ($\Delta F = F(r_{+f}) - F(r_{+i})$) and the quantum work $\left(e^{\frac{\Delta F}{T}}\right)$ \cite{Soroushfar:2023mrm} as follows:
\begin{align}\label{Delta-F}
\Delta F = -\frac{r_{+f}^{\eta^{2}} \pi^{1-\frac{\eta^{2}}{4}} {\mathrm e}^{-\frac{\pi  r_{+f}^{2}}{2}}WM \! \left({\frac{\eta^{2}}{4},\frac{\eta^{2}}{4}+\frac{1}{2}}, \pi  r_{+f}^{2}\right)   \left(r_{+f}^{2}\right)^{-\frac{\eta^{2}}{4}} \lambda}{\pi  \left(\eta^{2}+2\right)}-\\
\frac{\lambda  \left(\frac{\eta^{2} r_{+f}^{\eta^{2}-2}}{2}+r_{+f}^{\eta^{2}} \pi \right) {\mathrm e}^{-\pi  r_{+f}^{2}}+\frac{\pi  \left(\eta^{2}-2\right) \left(r_{+f}^{\eta^{2}}-r_{+i}^{\eta^{2}}\right)}{2}}{2 \pi},
\end{align}

and
\begin{equation}\label{W}
	W=e^{\frac{\Delta F}{T}}= {\mathrm e}^{-\frac{2 r_{+}^{-\eta^{2}+2} \left(\alpha1 +\alpha2 \right)}{\left(\eta^{2}+2\right) \eta^{2}}},
\end{equation}

by which

\begin{equation}\label{Delta S2}
	\alpha1= 2 r_{+f}^{\eta^{2}} \pi^{1-\frac{\eta^{2}}{4}} {\mathrm e}^{-\frac{\pi  r_{+f}^{2}}{2}}M_{W} \! \left({\frac{\eta^{2}}{4},\frac{\eta^{2}}{4}+\frac{1}{2}}, \pi  r_{+f}^{2}\right)   \left(r_{+f}^{2}\right)^{-\frac{\eta^{2}}{4}} \lambda,
\end{equation}

\begin{equation}\label{Delta S2}
	\alpha2=\left(\lambda  \left(\frac{\eta^{2} r_{+f}^{\eta^{2}-2}}{2}+r_{+f}^{\eta^{2}} \pi \right) {\mathrm e}^{-\pi  r_{+f}^{2}}+\frac{\pi  \left(\eta^{2}-2\right) \left(r_{+f}^{\eta^{2}}-r_{+i}^{\eta^{2}}\right)}{2}\right) \left(\eta^{2}+2\right).
\end{equation}
Equation (\ref{W}) can be used to express quantum work in terms of the partition functions, denoted as $W = \frac{Z_f}{Z_i}$. The quantum work ($e^{\frac{\Delta F}{T}}$) for various values of $ \lambda $ and $ \eta $ can be visualized in relation to the horizon radius $ r_{+} $, as shown in Fig. \ref{Pic:W}.

The distribution of partition functions for a black hole is influenced by the relative weights, or probabilities, associated with transitions between different states \cite{Pourhassan:2021mhb}. These weights are determined by the quantum work performed during the transition. Quantum work relies on the difference in equilibrium free energies between the initial and final states, which, in turn, depends on the microstates of the black hole. As the black hole emits radiation, understanding this process requires consideration of the average quantum work during the emission process \cite{Pourhassan:2022opb}. The significance of quantum work becomes more pronounced at small scales, necessitating the incorporation of quantum gravitational corrections. It is evident from the curves in Fig. \ref{Pic:W} that, at larger values of $r_{+}$, all curves coincide. These corrections modify equilibrium free energies, subsequently affecting the calculation of quantum work. Therefore, it becomes imperative to integrate these modified expressions for free energies to accurately assess the impact of quantum gravitational corrections on the distribution of quantum work \cite{Pourhassan:2022auo,Pourhassan:2020all,Pourhassan:2022opb,Pourhassan:2021mhb}.

\section{Conclusion}\label{con}
In a recent study, a novel solution has been introduced that involves a dirty/hairy black hole within the framework of ENE-dilaton theory\cite{Mazharimousavi:2023tbn}. This particular black hole is characterized by a uniform radial electric field and a singular dilaton scalar field, which is NAF and possesses a singularity at its core. This type of black hole is of significant interest due to its connections with charged black holes in string theory, where the hair/dilaton field is non-minimally coupled to electromagnetic fields. These connections have far-reaching implications in various areas, such as the AdS/CFT correspondence, which hints at a holographic duality with a quark-gluon plasma \cite{JHEP7}. Consequently, there is a pressing need to delve into the thermodynamics of this intriguing black hole, especially at quantum scales, which is the central focus of the present paper.

The paper has commenced by providing a comprehensive overview of the dirty black hole sustained by a uniform electric field in the ENE-dilaton theory. Subsequently, we have derived the entropy of this particular dirty black hole, accounting for exponential corrections. We have also conducted an in-depth examination of various thermodynamic properties associated with this black hole. An analysis of the heat capacity reveals that the final stage of this black hole becomes unstable, assuming that the correction parameter $\lambda$ is positive, as anticipated. Furthermore, we have computed the quantum work employing free energy and investigated the influence of exponential corrections on it. Remarkably, it is observed that both non-perturbative and perturbative corrections exert a substantial impact, particularly as the black hole size (horizon radius) diminishes due to the process of Hawking radiation.

In light of these findings, future research avenues emerge from this article. First and foremost, it is imperative to explore the consequences of the black hole's instability in the late stages of its evolution, which could have profound implications for our understanding of gravitational dynamics. Additionally, further investigations into the holographic duality between this type of black hole and quark-gluon plasma, as suggested by the AdS/CFT correspondence, could yield valuable insights into the behavior of matter under extreme conditions. Finally, the interplay between the dilaton field and thermodynamic properties of black holes remains a fertile ground for future research, particularly in the context of quantum gravity and string theory. These avenues promise exciting developments in our comprehension of the fundamental aspects of the universe.

\section {Acknowledgements}
We wish to convey our appreciation to Prof. Dr. S. H. Mazharimousavi for his insightful comments on Eqs. \eqref{2.5}-\eqref{2.7}, and enriching discussions regarding the topic of dirty black holes. \.{I}.S. would like to acknowledge networking support of COST Actions CA21106 and CA22113. He also thanks to T\"{U}B\.{I}TAK, SCOAP3, and ANKOS for their support.

\end{document}